# Inducing Chalcogenide Phase Change with Ultra-Narrow Carbon Nanotube Heaters

Feng Xiong,[1] Albert Liao,[1] and Eric Pop[1, 2, *]

[1]*Dept. of Electrical and Computer Engineering, Micro and Nanotechnology Laboratory, University of Illinois, Urbana-Champaign, IL 61801, USA*

[2]*Beckman Institute, University of Illinois, Urbana-Champaign, IL 61801, USA*

Carbon nanotube (CNT) heaters with sub-5 nm diameter induce highly localized phase change in $Ge_2Sb_2Te_5$ (GST) chalcogenide. A significant reduction in resistance of test structures is measured as the GST near the CNT heater crystallizes. Effective GST heating occurs at currents as low as 25 µA, significantly lower than in conventional phase change memory with metal electrodes (0.1–0.5 mA). Atomic force microscopy reveals nucleation sites associated with phase change in GST around the CNT heater. Finite element simulations confirm electrical characteristics consistent with the experiments, and reveal the current and phase distribution in GST.

[*]Contact: epop@illinois.edu

Phase change memory (PCM) is a promising candidate for next-generation non-volatile data storage,[1] combining features such as low operating voltage (~1 V), fast access times, high endurance, and potential for scaling below the size limits of Flash memory (~35 nm).[2] PCM devices rely on phase change materials like chalcogenides (e.g. $Ge_2Sb_2Te_5$ or GST) which exhibit sharply contrasting and switchable electrical resistivity and optical reflectivity[3] between their crystalline and amorphous states. PCM switching is induced through electrical (Joule) heating, which requires relatively high programming currents (0.1–0.5 mA) even when nanoscale heaters[2,4] or GST nanowires[5] are used. Reducing this programming current is one of the most significant challenges associated with PCM today.

In this study, we have successfully used individual carbon nanotubes (CNTs) as nanoscale heaters to induce ultra-narrow phase change regions in GST, while applying currents on the order of 10 μA. We found that GST sputtering is compatible with CNT devices, with conformal deposition and little apparent damage to the CNT. The low currents needed to induce phase change are a result of the excellent thermal stability (up to >1000 $^o$C) and extremely small diameter (<5 nm) of the CNT heaters. We have also performed atomic force microscopy (AFM) imaging to examine the phase-change regions, and implemented a 3-dimensional finite-element (FE) model to understand the resistive changes observed experimentally.

CNT devices used in this work are grown directly on $SiO_2$ and contacted with Pd electrodes as described in Ref. 6 and the EPAPS supplementary material at *[URL will be inserted by AIP]*. We obtain both single-wall and small diameter (<5 nm) multi-wall CNTs, and we find both can be used to induce phase change in GST. A 10 nm amorphous film of $Ge_2Sb_2Te_5$ (GST) was sputtered on top of the CNT devices, as shown in Fig. 1a. Atomic force microscope (AFM) measurements confirm the thin GST was conformal, and surface roughness was minimally increased from ~0.3 nm (bare $SiO_2$) to ~0.5 nm (after GST deposition), as shown in the SOM. Moreover, electrical measurements of the CNT before and immediately after GST sputtering (Fig. 1b) indicate only ~20% change in CNT resistance, suggesting little damage to the nanotube from the sputtering process.

After GST sputtering we performed several compliance-limited DC current sweeps, while monitoring the voltage across the device, as shown in Fig. 2. Although the thin GST film spans between the two electrodes, its amorphous resistivity is very high (~100 Ω·cm), and the current is entirely carried by the CNT during the initial sweep (labeled #1). Subsequent sweeps to higher currents (#2 – #6) lead to increasing conductivity with voltage snapback, attributed to a gradual transition of the GST surrounding the CNT from amorphous to crystalline. At higher currents the temperature of the CNT increases significantly,[7] and a low-resistance crystalline GST "sleeve" begins to form around the CNT. Once the phase transition occurs, the crystalline state of GST is preserved as seen from hysteresis loops in Fig. 2a, with each forward sweep following the previous backward sweep (e.g. black and magenta arrows between #2 and #3). Sweeps labeled #1 – #6 were made by gradually increasing the upper current limit in 20 μA increments. Consequently, the resistance of the CNT-GST structure is reduced by more than an order of magnitude, as an increasing volume of GST surrounding the CNT gradually heats up and crystallizes, introducing a parallel current flow path. Once the current reached ~160 μA (sweep #7) the GST was irreversibly damaged, but the measured *I-V* returned along the original path (#1), indicating the CNT itself was still conducting, unchanged, and undamaged. The last point highlights the resilience of CNTs even under the most extreme conditions, and their durability as nanoscale GST heaters. We note the heating current at which GST phase transition first occurs (~25 μA) is much lower than in conventional PCM, although voltages are higher due to the relatively long, resistive CNT (~400 kΩ). Shorter CNTs (<1 μm) with good contacts have resistance an order of magnitude lower,[8] and would yield effective heating at voltages that are proportionally decreased as well.[7]

AFM images in Figs. 2b–2e were taken after sweeps #1, 3, 6 and 7 respectively, showing the progression of the GST surface as the structure is pushed to higher currents. As the GST begins to crystallize, nucleation points form along the length of the heated CNT, consistent with observations in GST nanowires.[9] These nucleation centers eventually lead to GST volume changes, possible void formation (likely due to thermal expansion mismatch)[10] and delamination from the CNT as the height profile increases up to ~7 nm. The region of nucleation follows the well-known Joule heating temperature profile of a CNT on $SiO_2$.[7,11] In fact, Fig. 2c reveals that phase and volume changes in GST occur initially near the middle of the CNT, where the temperature is highest. This leads to a "visualization" of the thermal healing length ($L_H \sim 0.25$ μm) along the CNT, i.e. the length scale over which heat sinking from the metal contacts remains effective.[7,11] At distances greater than $L_H$ from the contacts, the CNT heat sinking is limited by the $SiO_2$ substrate. Eventually the entire GST around the CNT heats up and crystallizes (Fig. 2d), leading to the large measured increase in conductance. The GST breakdown electrically observed at ~160 μA appears as a physical "bubbling" and delamination of the thin film (Fig. 2e).

To better understand the heating and crystallization in this test structure, we implemented a 3-dimensional (3-D) finite element (FE) model[12] that self-consistently takes into account the electrical, thermal and Joule heating interactions. The simulated structure shown in Fig. 3a mimics that of the experimental devices (Fig. 1 and Fig. 2). In the electrical model, the Poisson and continuity equations are solved to obtain the voltage and current distribution in the device. Simulation parameters are similar to Ref. 13 and summarized in the SOM. In addition, the electrical[14] and thermal conductivity[15] of GST ($\sigma_{GST}$ and $k_{GST}$) are parameterized as a function of phase and temperature, as shown in the SOM. The transition temperatures are taken as those well-known for GST, i.e. 150 °C for the amorphous to fcc transition (~1000× $\sigma_{GST}$ increase), and 350 °C for the fcc to hcp transition (additional ~10× $\sigma_{GST}$ increase).

The electrical conductivity and Joule heating of the CNT are calculated following Ref. 7, including the temperature- and position-dependent carrier mean free paths. On all external boundaries, electrical insulation boundary conditions apply, except across the electrodes where a constant current flow is applied. The heat diffusion equation is used to obtain the 3-D temperature and phase distribution in the device. In other words, when the GST reaches a phase transition temperature, it is switched to the corresponding phase, which is then preserved upon return to room temperature (Fig. S2 in EPAPS supplementary material at *[URL will be inserted by AIP]*).

Adiabatic thermal boundary conditions are used on all exterior boundaries (convective air cooling and radiation loss are insignificant) except for the bottom boundary of the $SiO_2$/Si interface, where a constant temperature 300 K is assumed. At interior boundaries, thermal boundary resistance (TBR, $R_{th}$) is used to model the heat fluxes and temperature gradients at the interfaces.[16] The TBR is included by adding a thin thermally resistive layer with thickness $d_{th}$ and thermal conductivity $k_{th}$ such that $R_{th} = d_{th}/k_{th}$. The Pd/CNT boundary is assumed to have $R_{th,c} = 1.2 \times 10^7$ K/W, and a thermal boundary conductance $g = 0.17$ W/K/m per CNT length is applied at the CNT/$SiO_2$ boundary.[7,17] All other interior boundaries have $R_{th} = 2.5 \times 10^{-8}$ m$^2$K/W per unit area, and all TBR is assumed to be temperature independent.[13,18]

Figure 3b displays typical current-voltage characteristics computed with this model (line), compared to experimental data (circles) for a typical CNT/GST device with $L \approx 3$ μm, and $d = 3.2$ nm. The simulation performs current sweeps while monitoring the voltage, as does the experimental data. No changes are noted in the simulated *I-V* characteristics as the GST warms up beyond ~150 °C and changes into fcc crystalline state. At $I \approx 30$ μA the temperature in the GST surrounding the CNT heater reaches ~350 ˚C, the transition temperature of GST from fcc to hcp state. As more GST switches to the highly conductive hcp state, the resistance of the device begins to decrease significantly, with a parallel current path being created in the GST. Hence, the voltage decreases even as the current increases. Figures 3c and 3d show Y-Z plane cross-sections of temperature and electrical conductivity in GST at the center of the CNT before and after the voltage snapback seen in the simulated device $I$-$V$ curve. At $I = 35$ µA the GST directly above the CNT heater has partially switched to the highly conductive hcp state. At $I = 50$ µA, a significant amount of GST near the CNT was transformed into the hcp state. A parallel current path is now available in GST which causes the voltage snapback shown in Fig. 3b.

In the backward sweep, the hcp state of GST is preserved upon return to room temperature, and the $I$-$V$ curve follows a lower resistance path. Thus, interestingly, the simulations indicate that voltage snapback in this test structure is due to the fcc to hcp transition of GST, not the amorphous to fcc transition. This occurs although the resistance of GST decreases by three orders of magnitude in the fcc phase (from GΩ to MΩ), the total resistance of the entire device is still dominated by the CNT (~0.1 MΩ). Only a transition to the hcp state brings the GST "sleeve" surrounding the CNT into a resistance range comparable to that of the CNT, leading to a measurable change in the electrical characteristics (Figs. 2 and 3). As heating to the hcp phase is easily reached, the CNT heater can also be used to melt the GST,[19] although an optimized test structure should be employed to achieve fast (~ns) quenching into the amorphous phase.

In conclusion, we have demonstrated that CNT heaters with sub-5 nm diameters can induce highly localized phase change in GST thin films, with heating currents of the order ~25 µA. The current-voltage characteristics of simple test devices show voltage snapback behavior, indicating GST switching from highly resistive (amorphous) to highly conductive (hcp crystalline) states. Additional AFM characterization and 3-D finite-element modeling confirm the morphological and phase changes occurring. The simulation platform can also be used for future studies, to provide insight into optimized, lower current, and reversible switching device structures. This proof-of-concept study opens up the possibility of developing phase-change memory cells with carbon nanotube heaters and ultra-low switching energy.

We acknowledge support from the Office of Naval Research grant N00014-09-1-0180, the DARPA Young Faculty Award (YFA), and the NRI Coufal Fellowship to A.L. We are indebted to D. Estrada, K. Darmawikarta and Prof. J. Abelson for technical support.

## REFERENCES

[1] H. F. Hamann, M. O'Boyle, Y. C. Martin, M. Rooks, and K. Wickramasinghe, Nature Materials **5** (5), 383 (2006).

[2] S. Raoux, G.W. Burr, M.J. Breitwisch, C.T. Rettner, Y.-C. Chen, R.M. Shelby, M. Salinga, D. Krebs, S.-H. Chen, H.-L. Lung, and C.H. Lam, IBM J. Res. & Dev. **52** (4/5), 465 (2008).

[3] A. Redaelli, A. Pirovano, F. Pellizzer, A.L. Lacaita, D. Ielmini, and R. Bez, IEEE Elec. Dev. Lett. **25** (10), 684 (2004).

[4] I. V. Karpov and S. A. Kostylev, IEEE Elec. Dev. Lett. **27** (10), 808 (2006).

[5] S.-H. Lee, Y. Jung, and R. Agarwal, Nature Nanotechnology **2**, 626 (2007).

[6] A. Liao, Y. Zhao, and E. Pop, Physical Review Letters **101** (25) (2008).

[7] E. Pop, D. A. Mann, K. E. Goodson, and H. J. Dai, Journal of Applied Physics **101** (9) (2007).

[8] A. Javey, J. Guo, M. Paulsson, Q. Wang, D. Mann, M. Lundstrom, and H. J. Dai, Physical Review Letters **92** (10) (2004).

[9] Se-Ho Lee, Yeonwoong Jung, and Ritesh Agarwal, Nano Letters **8** (10), 3303 (2008); S. Meister, D. T. Schoen, M. A. Topinka, A. M. Minor, and Y. Cui, Nano Letters **8** (12), 4562 (2008).

[10] S. W. Nam, D. Lee, M. H. Kwon, D. M. Kang, C. Kim, T. Y. Lee, S. Heo, Y. W. Park, K. Lim, H. S. Lee, J. S. Wi, K. W. Yi, Y. Khang, and K. B. Kim, Electrochemical and Solid State Letters **12** (4), H155 (2009); Sung-Min Yoon, Kyu-Jeong Choi, Nam-Yeal Lee, Seung-Yun Lee, Young-Sam Park, and Byoung-Gon Yu, Applied Surface Science **254** (1), 316 (2007); Sung-Hoon Hong and Heon Lee, Japanese Journal of Applied Physics **47**, 3372 (2008).

[11] Li Shi, Jianhua Zhou, Philip Kim, Adrian Bachtold, Arun Majumdar, and Paul L. McEuen, Journal of Applied Physics **105** (10), 104306 (2009).

[12] COMSOL Multiphysics, http://www.comsol.com.

[13] I. R. Chen and E. Pop, IEEE Transactions on Electron Devices **56** (7), 1523 (2009).

[14] M. Lankhorst, B. Ketelaars, and R. Wolters, Nature Materials **4**, 347 (2005).

[15] H. K. Lyeo, D. G. Cahill, B. S. Lee, J. R. Abelson, M. H. Kwon, K. B. Kim, S. G. Bishop, and B. K. Cheong, Applied Physics Letters **89** (15) (2006); J.P. Reifenberg, M.A. Panzer, S. Kim, A.M. Gibby, Y. Zhang, S. Wong, H.-S. P. Wong, E. Pop, and K.E. Goodson, Appl. Phys. Lett. **91**, 111904 (2007).

[16] J.P. Reifenberg, D.L. Kencke, and K.E. Goodson, IEEE Elec. Dev. Lett. **29** (10), 1112 (2008).

[17] E. Pop, Nanotechnology **19**, 295202 (2008).

[18] D. G. Cahill, W. K. Ford, K. E. Goodson, G. D. Mahan, A. Majumdar, H. J. Maris, R. Merlin, and S. R. Phillpot, Journal of Applied Physics **93** (2), 793 (2003).

[19] Probably in part responsible for the delamination of the GST at high power in Fig. 2e.

**FIGURES**

**Single-column positioning:**

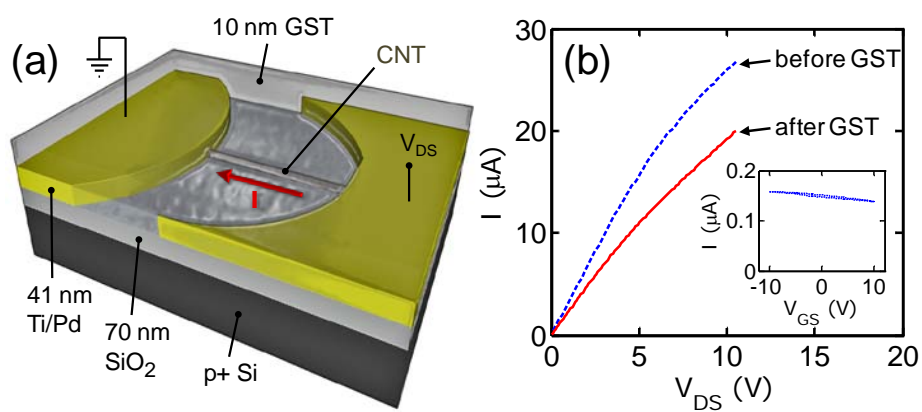

**FIG. 1:** (a) Schematic of CNT test structure with GST thin film sputtered on top. (b) Measured current-voltage of a typical CNT ($L \approx 1.88$ μm, $d \approx 3.3$ nm) before and after GST deposition. The inset displays the measured current vs. back-gate voltage, indicating metallic behavior.

**Single-column positioning:**

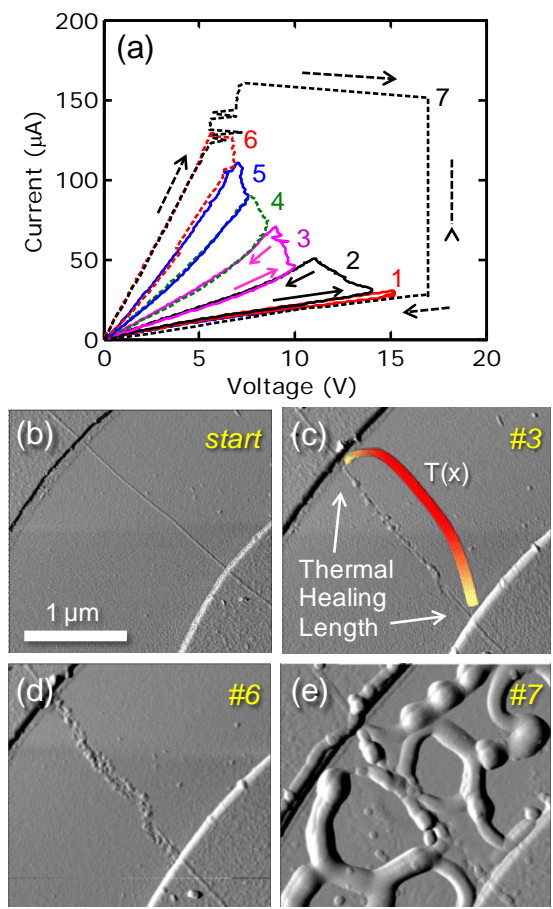

**FIG. 2:** Measured and imaged phase change of GST covering CNT heater. (a) Consecutive current sweeps to progressively higher current. Each state is preserved when the GST is returned to zero current (room temperature). (b) Topographic AFM before any current is applied and (c-e) after current sweep to ~70, 130, and 160 μA respectively. The latter correspond to traces #3, 6 and 7 labeled in (a). As current passes through the device, the CNT heats up and crystallizes the surrounding GST. In (c), the color profile shows the qualitative Joule heating temperature rise of the CNT, e.g. see Refs. 7 and 11. The GST near the middle of the CNT is crystallized first, illustrating the role of heat sinking at the CNT contacts (thermal healing length ~0.25 μm). At higher currents the GST covering the entire CNT is crystallized as shown in (d), and eventually fails (e).

**Two-column positioning:**

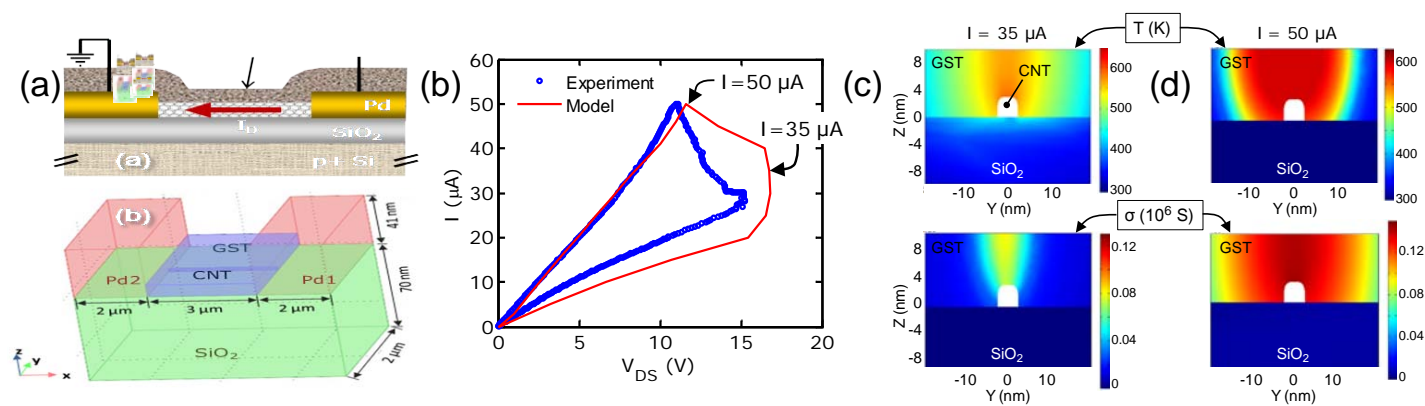

**FIG. 3:** Simulation of GST heating and phase change with CNT heater. (a) Device structure used in the 3-D simulation mimics the experimental test structures. (b) Measured (symbols) and simulated (lines) *I-V* characteristics of a typical metallic CNT covered by GST. The FE model predicts a sudden increase in overall conductivity as the GST changes to hcp crystalline state. (c) and (d) Cross-sectional temperature (top) and conductivity (bottom) of GST at the middle of the CNT at *I* = 35 μA and 50 μA, respectively. The simulations suggest that noticeable voltage snapback only occurs when ~5-10 nm of GST near the CNT transitions to hcp phase.